\documentclass[aps,prd,showpacs,preprint]{revtex4}
\usepackage{amsmath}
\usepackage{amssymb}

\begin{document}

\title{Bounds on the basic physical parameters for anisotropic compact
       general relativistic objects}
\author{C. G. B\"ohmer}
\email{boehmer@hep.itp.tuwien.ac.at} \affiliation{ASGBG/CIU,
Department of Mathematics, Apartado Postal C-600, University of
             Zacatecas (UAZ), Zacatecas, Zac 98060, Mexico}
\affiliation{Department of Physics, The University of Hong Kong,
Pokfulam Road, Hong Kong SAR, P. R. China}

\author{T. Harko}
\email{harko@hkucc.hku.hk} \affiliation{Department of Physics and
Center for Theoretical and Computational Physics, The University
of Hong Kong, Pokfulam Road, Hong Kong SAR, P. R. China}
\date{\today }

\begin{abstract}

We derive upper and lower limits for the basic physical parameters
(mass-radius ratio, anisotropy, redshift and total energy) for
arbitrary anisotropic general relativistic matter distributions in
the presence of a cosmological constant. The values of these
quantities are strongly dependent on the value of the anisotropy
parameter (the difference between the tangential and radial
pressure) at the surface of the star. In the presence of the
cosmological constant, a minimum mass configuration with given
anisotropy does exist. Anisotropic compact stellar type objects
can be much more compact than the isotropic ones, and their radii
may be close to their corresponding Schwarzschild radii. Upper
bounds for the anisotropy parameter are also obtained from the
analysis of the curvature invariants. General restrictions for the
redshift and the total energy (including the gravitational
contribution) for anisotropic stars are obtained in terms of the
anisotropy parameter. Values of the surface redshift parameter
greater than two could be the main observational signature for
anisotropic stellar type objects.
\end{abstract}

\pacs{04.40.Dg, 97.10.-q, 04.20.-q}

\maketitle

\section{Introduction}

It is generally believed that for smooth equations of state no
stable stellar configurations with central densities above that
corresponding to the limiting mass of neutron stars is stable
against acoustical vibrational modes \cite{1}. The maximum allowed
gravitational mass of a neutron star has been derived by using the
properties of neutron matter at density ranges where they can be
accurately predicted and imposing a minimum number of constraints
at densities exceeding a higher fiducial density, $\rho _{0}$,
e.g. subluminal sound velocity and thermodynamic stability .
Following this approach it has been rigorously proved that the
mass of a stable neutron star becomes maximum for the stiffest
possible equation of state that is consistent with the fundamental
physical constraints \cite{2}. As a result a maximum neutron star
mass of $3.2M_{\odot }$ has been found. On the other hand, by
using the static spherically symmetric gravitational field
equations, Buchdahl \cite{3} has obtained an absolute constraint
of the maximally allowable mass $M$-radius $R$ ratio for isotropic
fluid spheres
of the form $2M/R<8/9$ (in the present paper we use natural units so that $%
c=G=1$).

The study of the maximum mass and mass-radius ratio for compact
stars has been done mainly for isotropic stellar objects, in which
the tangential pressure equals the radial one. But, as suggested
by Ruderman \cite{4}, theoretical investigations of more realistic
stellar models show that the stellar matter may be anisotropic at
least in certain very high density ranges ($\rho >10^{15}$
g/cm$^{3}$), where the nuclear interactions must be treated
relativistically. According to these views in such massive stellar
objects the radial pressure $p_{r}$ may not be equal to the tangential one $%
p_{\perp }$, $p_{r}\neq p_{\perp }$. No celestial body is composed
of purely perfect fluid. Anisotropy in fluid pressure could be
introduced by the existence of a solid core or by the presence of
type 3A superfluid \cite{5}, different kinds of phase transitions
\cite{6}, pion condensation \cite{7} or by other physical
phenomena. A slowly rotating system can be formally described as a
static anisotropic fluid \cite{8}. The mixture of two gases (e.g.,
monatomic and molecular hydrogen, or ionized hydrogen and
electrons) can also be interpreted as an anisotropic fluid
\cite{9}. For a review of the appearance of local anisotropy in
self-gravitating systems and of its main physical consequences see
\cite{HeSa97}.

For arbitrarily large anisotropy, in principle there is neither
limiting mass nor limiting redshift \cite{10}. Semi-realistic
equations of state lead to a mass of $3-4M_{\odot }$ for neutron
stars with an anisotropic equation of state \cite{10}. Bowers and
Liang \cite{11} have analytically obtained the maximum equilibrium
mass and surface redshift in the case of incompressible neutron
matter. They also numerically investigated models with a special
form of anisotropy, founding that specific models lead to
increases in the redshift proportional to the deviations from
isotropy.

Bondi \cite{12} considered the relation between redshift and the
ratio of the trace of the pressure tensor to local density. When
anisotropic pressures are allowed considerably larger redshift
values can be obtained. Several classes of solutions of the
gravitational field equations for anisotropic matter distributions
have been obtained in \cite{HaMa02} and \cite{MaHa03}.

The value of the bound $M/R$ is an important problem in
relativistic astrophysics since ``the existence of such a bound is
intriguing because it occurs well before the appearance of an
apparent horizon at $M=R/2$'' \cite{13}. In \cite{13} the upper
limit of $M/R$ for compact general relativistic configurations has
been re-investigated by assuming that inside the star the radial
stress $p_{r}$ is different from the tangential one $p_{\perp }$.
If the density is monotonically decreasing and $p_{r}\geq p_{\perp
}$ then the upper bound $8/9$ is still valid to the entire bulk if
$m$ is replaced by
the quasi-local mass. This bound cannot be recovered if $p_{\perp }\geq $ $%
p_{r}$ and / or the density is not a monotonic function.

The maximum value of the redshift for anisotropic stars was
derived in \cite{Iv02}. For realistic anisotropic star models the
surface redshift cannot exceed the values $3.842$ or $5.211$ when
the tangential pressure satisfies the strong or the dominant
energy condition, respectively. Both values are higher than 2, the
bound in the perfect fluid case. Several bounds on the important
physical parameters for the anisotropic stars have been derived in
\cite{BaHaGl03}. If the radial pressure is larger than the
tangential pressure, then the radial pressure is also larger than
the corresponding pressure for a fiducial isotropic model with the
same mass function and total mass, while the opposite holds if the
tangential pressure is larger than the radial one. By imposing an
energy condition, the value for the maximum possible redshift at
the surface of the star can be obtained.

Several bounds on the mass-radius ratio and anisotropy parameter
have also been found, for models in which the anisotropy increases
as $r^2$, in \cite{MaDoHa02}.

One of the most important results of modern cosmology is the
observational evidence for the existence of the cosmological
constant. The first pressing piece of data involved the study of
Type Ia Supernovae. Observations of Type Ia Supernovae with
redshift up to about $z\sim 1$ provided evidence that we may live
in a low mass-density Universe, with the contribution of the
non-relativistic matter (baryonic plus dark) to the total energy
density of the Universe of the order of $\Omega _{m}\sim 0.3$
\cite{Ri98,Pe98}. The value of $\Omega _{m}$ is significantly less
than unity \cite{OsSt95}, and consequently either the Universe is
open or there is some additional energy density $\rho $ sufficient
to reach the value $\Omega_{\mathrm{total}}=1$, predicted by
inflationary theory.

The existence of the cosmological constant modifies the allowed
ranges for various physical parameters, like, for example, the
maximum mass of compact stellar objects, thus leading to
modifications of the ``classical'' Buchdahl limit \cite{MaDoHa00}.

On the other hand, we cannot exclude \textit{a priori} the
possibility that the cosmological constant, as a manifestation of
vacuum energy, may play an important role not only at galactic or
cosmological scales, but also at the level of elementary
particles. With the use of the generalized Buchdahl identity
\cite{MaDoHa00}, it can be rigorously proven that the existence of
a non-negative $\Lambda $ imposes a lower bound on the mass $M$ and density $%
\rho $ of general relativistic objects of radius $R$, which is
given by \cite{BoHa05}
\begin{equation}
2M\geq \frac{8\pi \Lambda }{6}R^{3},\qquad \rho =\frac{3M}{4\pi
R^{3}}\geq \frac{\Lambda }{2}=:\rho _{\min }.  \label{minm}
\end{equation}

Therefore, the existence of the cosmological constant implies the
existence of an absolute minimum mass and density in the universe.
No object present in relativity can have a density that is smaller
than $\rho_{\min }$. For $\Lambda >0$ this result also implies a
minimum density for stable fluctuations in energy density.

There are some other astrophysical systems which may be modeled,
at least at a qualitative level, by using an effective
``cosmological constant''. For example, there is the possibility
that scalar fields present in the early universe could condense to
form the so named boson stars \cite{Ry97,MiSc00,ScMi03}. The
simplest kind of boson star is made up of a self-interacting
complex scalar field $\Phi $ describing a state of zero
temperature \cite{ScMi03,SeSu90,Ku91}. If we suppose that in the
star's interior regions and for some field configurations $\Phi$
is a slowly varying function of $r$, so that it is nearly a
constant, then in the gravitational field equations the scalar
field will play the role of a cosmological constant, which could
also describe a mixture of ordinary matter and bosonic particles.

It is the purpose of the present paper to obtain the minimum and
the maximum allowable mass-radius ratio in the case of anisotropic
compact general relativistic objects, in the presence of a
cosmological constant, as a function of the anisotropy parameter.
We found that for anisotropic compact general relativistic bodies
an upper limit (different from half) does not exist in general.
Consequently there are no limits on the red shift of the radiation
coming from this type of objects. On the other the presence of the
anisotropy induces a minimum mass-radius ratio even in the absence
of the cosmological constant. Upper bounds on the anisotropy are
also derived by using the properties of the linear and quadratic
scalars formed out of the curvature tensor (the Ricci invariants).

The present paper is organized as follows. The Buchdahl limit for
anisotropic general relativistic objects in the presence of a
cosmological constant is derived in Section II. In Section III we
consider the problem of the minimum mass of anisotropic compact
objects. Upper limits for the surface anisotropy of relativistic
stars are obtained in Section IV. We discuss our results and
conclude our paper in Section V.

\section{The Buchdahl limit for anisotropic stars}

For a static general relativistic spherically symmetric matter
configuration the interior line element is given by
\begin{equation}
ds^{2}=e^{\nu }dt^{2}-e^{\lambda }dr^{2}-r^{2}\left (d\theta
^{2}+\sin ^{2}\theta d\varphi ^{2}\right ).
\end{equation}

We assume that the star consists of an anisotropic fluid
distribution in the presence of a cosmological constant. For such
a system the components of the energy-momentum tensor are
\begin{equation}
T_{0}^{0}=\rho +\Lambda ,\quad T_{1}^{1}=-p_{r}+\Lambda ,\quad
T_{2}^{2}=T_{3}^{3}=-p_{\perp }+\Lambda ,
\end{equation}
where $\rho $ is the energy density and $\Lambda $ is the
cosmological constant.

We suppose that inside the star $p_{r}\neq p_{\perp }$, $\forall
r\neq 0$. We define the anisotropy parameter as $\Delta =p_{\perp
}-p_{r}$. $\Delta $ is a measure of the deviations from isotropy.
If $\Delta >0,\forall r\neq 0$
the body is tangential pressure dominated while $\Delta <0$ indicates that $%
p_{r}>p_{\perp }$. In realistic physical models for compact stars,
$\Delta $ should be finite, positive and should satisfy the
dominant energy condition (DEC) $\Delta \leq \rho $ and the strong
energy condition (SEC) $2\Delta +p_{r}\leq \rho $. These
conditions may be written together as $\Delta \leq n\rho $, where
$n=1$ for DEC and $n=1/2$ for SEC, if the realistic condition for
the positivity of $p_{r}$ in the interior is accepted \cite{Iv02}.

The properties of the anisotropic compact object can be completely
described by the gravitational structure equations, which are
given by:
\begin{equation}  \label{5}
\frac{dm}{dr}=4\pi \rho r^{2},
\end{equation}
\begin{equation}  \label{6}
\frac{dp_{r}}{dr}=-\frac{\left (\rho +p_{r}\right )\left [m+4\pi
r^{3}\left
(p_{r}-\frac{2\Lambda }{3}\right )\right ]}{r^{2}\left (1-\frac{2m}{r}-\frac{%
8\pi }{3}\Lambda r^2\right )}+\frac{2\Delta }{r},
\end{equation}
\begin{equation}  \label{7}
\frac{d\nu }{dr}=-\frac{2}{\rho +p_{r}}\frac{dp_{r}}{dr}+\frac{4\Delta }{%
r\left (\rho +p_{r}\right )},
\end{equation}
where $m(r)$ is the mass inside radius $r$ .

In the Newtonian limit and in the absence of the cosmological constant Eq.~(%
\ref{6}) reduces to the expression \cite{HeSa97}
\begin{equation}
\frac{dp_{r}}{dr}=-\frac{m\rho }{r^{2}}+\frac{2\Delta }{r}.
\end{equation}

Hence the anisotropy term is Newtonian in origin \cite{11}. A
solution of Eqs.~(\ref{5})-(\ref{7}) is possible only when
boundary conditions have been imposed. As in the isotropic case we
require that the interior of any matter
distribution be free of singularities, which imposes the condition $%
m(r)\rightarrow 0$ as $r\rightarrow 0$. Assuming that $p_{r}$ is finite at $%
r=0$, we have $\nu ^{\prime }\rightarrow 0$ as $r\rightarrow 0$.
Therefore the gradient $dp_{r}/dr$ will be finite at $r=0$ only if
$\Delta $ vanishes at least as rapidly as $r$ when $r\rightarrow
0$. This requires that the anisotropy parameter satisfies the
boundary condition
\begin{equation}
\lim _{r\rightarrow 0}\frac{\Delta \left (r\right )}{r}=0.
\end{equation}

At the center of the star the other boundary conditions for Eqs.~(\ref{5})-(%
\ref{7}) are $p_{r}(0)=p_{\perp }(0)=p_{c}$ and $\rho (0)=\rho _{c}$, where $%
\rho _{c}$ and $p_{c}$ are the central density and pressure,
respectively.
The radius $R$ of the star is determined by the boundary condition $%
p_r\left (R\right )=0$. We do not necessarily require that the
tangential pressure $p_{\perp }$ vanishes for $r=R$. Therefore at
the surface of the star the anisotropy parameter satisfies the
boundary condition $\Delta (R)=p_{\perp }(R)-p_{r}(R)=p_{\perp
}(R)\geq 0$. To close the field equations the equations of state
of the radial pressure $p_r=p_r\left (\rho \right )$ and of the
tangential pressure $p_{\perp }=p_{\perp }\left (\rho \right )$
must also be given.

With the use of Eqs.~(\ref{5})-(\ref{7}) it is easy to show that
the function $\zeta =e^{\nu /2}>0,\forall r\in \lbrack 0,R\rbrack
$, obeys the equation
\begin{equation}
\label{9} \frac{y}{r}\frac{d}{dr}\left [\frac{y}{r}\frac{d\zeta
}{dr}\right ]=\frac{\zeta }{r}\left
[\frac{d}{dr}\frac{m(r)}{r^{3}}+\frac{8\pi \Delta }{r}\right ],
\end{equation}
where we denoted
\begin{equation}
\alpha (r)=1+\frac{4\pi }{3}\Lambda \frac{r^3}{m(r)}, \qquad y(r)
= \sqrt{1-\frac{2\alpha(r)m(r)}{r}}.
\end{equation}

For $\Delta =0$ and $\Lambda =0$ Eq.~(\ref{9}) reduces to the
isotropic equation considered in \cite{14}. Since the density
$\rho $ does not increase with increasing $r$, the mean density of
the matter $\langle \rho \rangle =3m(r)/4\pi r^{3}$ inside radius
$r$ does not increase either.

Therefore we assume that inside a compact general relativistic
object the condition
\begin{equation}
\frac{d}{dr}\frac{m(r)}{r^{3}}<0,
\end{equation}
holds independently of the equation of state of dense matter. By
defining a new function
\begin{equation}  \label{10}
\eta(r) = 8\pi \int _{0}^{r} \frac{r'}{y(r')} \left \{ \int
_{0}^{r'} \frac{\Delta(r'')}{y(r'')} \frac{\zeta(r'')}{r''} dr''
\right \}dr',
\end{equation}
denoting
\begin{equation}
\Psi =\zeta -\eta ,
\end{equation}
and introducing a new independent variable
\begin{equation}
\xi =\int _{0}^{r} \frac{r'}{y(r')} dr',
\end{equation}
from Eq.~(\ref{9}) we obtain the basic result that all stellar
type general relativistic matter distributions with negative
density gradient obey the condition
\begin{equation}  \label{12}
\frac{d^{2}\Psi }{d\xi ^{2}}<0,\quad \forall r\in \left [0,R\right
].
\end{equation}

Using the mean value theorem we conclude \cite{14}
\begin{equation}
\frac{d\Psi }{d\xi }\leq \frac{\Psi \left (\xi \right )-\Psi
(0)}{\xi },
\end{equation}
or, taking into account that $\Psi (0)>0$, we find
\begin{equation}
\Psi ^{-1}\frac{d\Psi }{d\xi }\leq \frac{1}{\xi }.
\end{equation}

In the initial variables we have
\begin{multline}  \label{15}
\frac{y(r)}{r} \left(\frac{1}{2}\frac{d\nu }{dr}e^{\nu (r)/2}-8\pi
\frac{r}{y(r)} \int _{0}^{r}\frac{\Delta(r') e^{\nu(r')/2}}{y(r')
r'} dr' \right)\\ \leq \frac{e^{\nu (r)/2}-8\pi \int _{0}^{r}
\frac{r'}{y(r')} \left( \int _{0}^{r'} \frac{\Delta \left
(r''\right )e^{\nu(r'')/2}}{y(r'') r''} dr'' \right) dr'}{\int
_{0}^{r} \frac{r'}{y(r')} dr'}.
\end{multline}

Since for stable stellar type compact objects $m/r^{3}$ does not
increase outwards, the condition
\begin{equation}
\frac{m(r^{\prime })}{r^{\prime }}\geq \frac{m(r)}{r}\left (\frac{r^{\prime }%
}{r}\right )^{2},\quad \forall r^{\prime }\leq r,
\end{equation}
holds for all points inside the star \cite{14}. Moreover, we
assume that in the presence of a cosmological constant, the
condition
\begin{equation}
\frac{\alpha \left (r^{\prime }\right )m\left (r^{\prime }\right )}{%
r^{\prime }}\geq \frac{\alpha \left (r\right )m\left (r\right )}{r}\left (%
\frac{r^{\prime }}{r}\right )^{2},
\end{equation}
or, equivalently,
\begin{equation}
\left [1+\frac{4\pi }{3}\Lambda \frac{r^{\prime 3}}{m\left
(r^{\prime }\right )}\right ]\frac{m\left (r^{\prime }\right
)}{r^{\prime }}\geq \left
[1+\frac{4\pi }{3}\Lambda \frac{r^{3}}{m\left (r\right )}\right ]\frac{%
m\left (r\right )}{r}\left (\frac{r^{\prime }}{r}\right )^{2},
\label{cond1}
\end{equation}
holds inside the compact object. In fact Eq.~(\ref{cond1}) is
satisfied for all values of the cosmological constant $\Lambda $
and is valid for all decreasing density compact matter
distributions.

In the following we assume that the anisotropy function satisfies
the general condition
\begin{equation}  \label{16}
\frac{\Delta (r^{\prime \prime })e^{\frac{\nu \left (r^{\prime
\prime
}\right )}{2}}}{r^{\prime \prime }}\geq \frac{\Delta (r^{\prime })e^{\frac{%
\nu \left (r^{\prime }\right )}{2}}}{r^{\prime }}\geq \frac{\Delta (r)e^{%
\frac{\nu (r)}{2}}}{r},r^{\prime \prime }\leq r^{\prime }\leq r.
\end{equation}

This condition is quite natural, taking into account that, since
the matter satisfies the dominant and strong energy conditions,
the anisotropy parameter is a monotonically decreasing function
inside the star.

Therefore we can evaluate the denominator in the RHS of
Eq.~(\ref{15}) as follows:
\begin{eqnarray}  \label{17}
\int _{0}^{r} \frac{r'}{y(r')}dr^{\prime }\geq
\int _{0}^{r}r^{\prime }\left [1-\frac{2\alpha \left (r\right )m(r)}{r^{3}}%
r^{\prime 2}\right ]^{-1/2}dr^{\prime }=\frac{r^{3}}{2\alpha \left
(r\right )m(r)}\left(1-y(r)\right).
\end{eqnarray}

For the second term in the bracket of the LHS of Eq.~(\ref{15}) we
find:
\begin{multline}  \label{19}
\int _{0}^{r} \frac{\Delta \left (r'\right )e^{\nu(r')/2}}{y(r')
r'}dr' \geq
\frac{\Delta(r)e^{\nu (r)/2}}{r}\int _{0}^{r}\left [1-\frac{2\alpha \left (r\right )m(r)%
}{r^{3}}r^{\prime 2}\right ]^{-1/2}dr^{\prime } \\ = \Delta
(r)e^{\nu (r)/2}\left [\frac{2\alpha \left (r\right
)m(r)}{r}\right
]^{-1/2} \arcsin\left (\sqrt{\frac{2\alpha \left (r\right )m(r)}{r}}%
\right ).
\end{multline}

The second term in the nominator of the RHS of Eq.~(\ref{15})
gives:
\begin{multline}
\label{20} \int _{0}^{r}\frac{r'}{y(r')} \left
\{\int_{0}^{r'}\frac{\Delta(r'')
e^{\nu(r'')/2}}{y(r'')r''}dr''\right \}dr' \\ \geq \int
_{0}^{r}r^{\prime 2}\frac{\Delta \left (r^{\prime
}\right )e^{\nu \left (y(r')r^{\prime }\right )/2}}{r^{\prime }}\left [\frac{%
2\alpha \left (r^{\prime }\right )m(r^{\prime })}{r^{\prime
}}\right ]^{-1/2} \arcsin \left (\sqrt{\frac{2\alpha \left
(r^{\prime }\right )m(r^{\prime })}{r^{\prime }}}\right
)dr^{\prime } \\ \geq \frac{\Delta (r)e^{\nu (r)/2}}{r}\int
_{0}^{r}r^{\prime 2} \left [1-\frac{2\alpha \left (r\right
)m(r)}{r^{3}}r^{\prime 2}\biggl/ \frac{2\alpha \left (r\right
)m(r)}{r^{3}}r^{\prime 2}\right ]^{-1/2} \arcsin\left
[\sqrt{\frac{2\alpha \left (r\right )m(r)}{r^{3}}}r^{\prime}\right
] dr^{\prime } \\ =
\Delta (r)e^{\nu (r)/2}r^{2}\left [\frac{2\alpha \left (r\right )m(r)}{r}%
\right ]^{-3/2}\left \{\sqrt{\frac{2\alpha \left (r\right
)m(r)}{r}}- y(r) \arcsin\left [\sqrt{\frac{2\alpha \left (r\right
)m(r)}{r}}\right ]\right \}.
\end{multline}

In order to obtain (\ref{20}) we have also used the property of
monotonic increase of the function $\arcsin x/x$ for $x\in \left
[0,1\right ]$.

Using Eqs.~(\ref{17})-(\ref{20}), Eq.~(\ref{15}) becomes:
\begin{multline}  \label{21}
\left \{1-\left [1-\frac{2\alpha \left (r\right )m(r)}{r}\right
]^{1/2}\right \}\frac{m(r)+4\pi r^{3}\left (p_{r}-\frac{2\Lambda }{3}\right )%
}{r^{3}\sqrt{1-\frac{2\alpha \left (r\right )m(r)}{r}}} \\ \leq
\frac{2\alpha \left (r\right )m(r)}{r^{3}}+8\pi \Delta (r)\left
\{\frac{
\arcsin\left [\sqrt{\frac{2\alpha \left (r\right )m(r)}{r}}\right ]}{%
\sqrt{\frac{2\alpha \left (r\right )m(r)}{r}}}-1\right \}.
\end{multline}

Eq.~(\ref{21}) is valid for all $r$ inside the star. It does not
depend on the sign of $\Delta $.

Consider first the isotropic case $\Delta =0$ and $\Lambda =0$. By
evaluating (\ref{21}) for $r=R$ we obtain
\begin{equation}
\frac{1}{\sqrt{1-\frac{2M}{R}}}\leq 2\left [1-\left (1-\frac{2M}{R}\right )^{%
\frac{1}{2}}\right ]^{-1},
\end{equation}
leading to the well-known result $2M/R\leq 8/9$ \cite{3},
\cite{14}.

By taking $\Delta =0$ but considering $\Lambda \neq 0$ we obtain
the following upper limit for the mass-radius ratio of a compact
object \cite{MaDoHa00}:
\begin{equation}
\frac{2M}{R}\leq \left (1-\frac{8\pi }{3}\Lambda R^{2}\right )\left [1-\frac{%
1}{9}\frac{\left (1-2\Lambda /\langle\rho\rangle \right )^{2}%
}{1-\frac{8\pi }{3}\Lambda R^{2}}\right ].
\end{equation}

Next consider the case $\Delta \neq 0$ and $\Lambda \neq 0$. We
denote
\begin{equation}
f\left (M,R,\Lambda ,\Delta \right )=2\frac{\Delta \left (R\right
)}{
\langle \rho \rangle }\left \{\frac{ \arcsin\left [\sqrt{\frac{%
2\alpha \left (R\right )M}{R}}\right ]}{\sqrt{\frac{2\alpha \left (R\right )M%
}{R}}}-1\right \}.
\end{equation}

Then Eq.~(\ref{21}) leads to the following restriction on the
mass-radius ratio for compact anisotropic stars in the presence of
a cosmological constant:
\begin{equation}  \label{24}
\frac{2M}{R}\leq \left (1-\frac{8\pi }{3}\Lambda R^{2}\right )\left [1-\frac{%
1}{9}\frac{\left (1-2\Lambda /\langle \rho \rangle \right )^{2}%
}{\left (1-\frac{8\pi }{3}\Lambda R^{2}\right )\left (1+f\right
)^{2}}\right].
\end{equation}

For a static general relativistic object the condition
$1-2M/R-8\pi \Lambda
R^2/3\geq 0$ must hold for all $R$, $M$ and $\Lambda $. Therefore from Eq.~(%
\ref{24}) we obtain that $\Delta (R)$ must obey the general
condition $(1-2\Lambda /\langle \rho \rangle)^{2}/(1+f)^2>0$,
which holds for all $\Delta $. Hence generally we cannot obtain
any limiting value for $\Delta (R)$ from Eq.~(\ref{24}). But for a
monotonically decreasing anisotropy several upper bounds for the
anisotropy parameter can be derived, as will be shown in Section
IV.

\section{The minimum mass of the anisotropic general relativistic objects}

On the vacuum boundary of the anisotropic star, corresponding to $r=R$, Eq.~(%
\ref{21}) takes the equivalent form
\begin{equation}  \label{an1}
\sqrt{1-\frac{2M}{R}-\frac{8\pi }{3}\Lambda R^{2}}\geq \frac{1}{3}\left (1-%
\frac{2\Lambda }{\langle \rho \rangle }\right )\frac{1}{1+f\left
(M,R,\Delta ,\Lambda \right )}.
\end{equation}

For small values of the argument the function $\arcsin x/x-1$
which appears in the definition of $f$ can be approximated as
$\arcsin x/x-1\approx x^{2}/6$. Therefore, Eq.~(\ref{an1}) can be
written as
\begin{equation}  \label{an2}
\sqrt{1-\frac{2M}{R}-\frac{8\pi }{3}\Lambda R^{2}}\geq \frac{M-\frac{8\pi }{3%
}\Lambda R^{3}}{3M+\frac{4\pi }{3}\Delta (R)R^{2}\left (2M+\frac{8\pi }{3}%
\Lambda R^3\right )}.
\end{equation}

By introducing a new variable $u$ defined as
\begin{equation}
u=\frac{M}{R}+\frac{4\pi }{3}\Lambda R^{2},
\end{equation}

Eq.~(\ref{an2}) takes the form
\begin{equation}
\sqrt{1-2u}\geq \frac{u-a}{bu-a},  \label{an3}
\end{equation}
where we denoted $a=4\pi \Lambda R^{2}$ and $b=3+8\pi \Delta \left
(R\right )R^{2}/3$, respectively. Then, by squaring we can
reformulate the condition given by Eq.~(\ref{an3}) as
\begin{equation}
u\left [2b^{2}u^{2}-\left (b^{2}+4ab-1\right )u+2a\left
(a+b-1\right )\right ]\leq 0,
\end{equation}
or, equivalently,
\begin{equation}  \label{cond2}
u\left (u-u_{1}\right )\left (u-u_{2}\right )\leq 0,
\end{equation}
where
\begin{equation}
u_{1}=\frac{b^{2}+4ab-1-\left (1-b\right
)\sqrt{(1+b)^{2}-8ab}}{4b^{2}},
\end{equation}
and
\begin{equation}
u_{2}=\frac{b^{2}+4ab-1+\left (1-b\right
)\sqrt{(1+b)^{2}-8ab}}{4b^{2}},
\end{equation}
respectively.

In the following we keep only the first order terms in both $\Lambda $ and $%
\Delta $. Since $u\geq 0$, Eq.~(\ref{cond2}) is satisfied if
$u\leq u_{1}$ and $u\geq u_{2}$, or $u\geq u_{1}$ and $u\leq
u_{2}$. However, the condition $u\geq u_{1}$ contradicts the upper
bound given by Eq.~(\ref{24}).

Therefore, Eq.~(\ref{cond2}) is satisfied if and only if for all
values of the physical parameters the condition $u\geq u_{2}$
holds. This is equivalent to the existence of a minimum bound for
the mass-radius ratio of compact anisotropic objects, which is
given by
\begin{equation}
u\geq \frac{2a}{1+b},
\end{equation}%
and explicitly written out using $a,b$ and $u$ as defined above
yields
\begin{equation}
\frac{2M}{R}\geq \frac{8\pi \Lambda }{6}R^{2}\left( \frac{1-\frac{4\pi }{3}%
\Delta R^{2}}{1+\frac{2\pi }{3}\Delta R^{2}}\right) .  \label{min}
\end{equation}

The presence of the anisotropy weakens the lower bound on the
mass, however, there still exists an absolute minimal mass in
nature. In the case $\Delta \equiv 0$, we recover the lower bound
for the minimum mass and density for isotropic general
relativistic objects, obtained in \cite{BoHa05}. In this case the
existence of a minimum mass is determined by the presence of the
cosmological constant only. For $\Lambda \equiv 0$, the presence
of an anisotropic pressure distribution reduces to the requirement
of the positivity of $M$, $M\geq 0$.

For isotropic systems with $\Delta \equiv 0$ and for a value of
the
cosmological constant of the order of $\Lambda \approx 3\times 10^{-56}$ cm$%
^{-2}$, the numerical value of minimum density following from Eq.~(\ref{min}%
) is $\rho _{\min }\approx 8\times 10^{-30}$ g/cm$^{3}$. By
assuming the existence in nature of an absolute minimum length of
the order of the Planck length $l_{Pl}$ it follows that the
corresponding absolute minimum mass is of the order of $1.4\times
10^{-127}$ g. However, by combining the minimum mass condition
with energy stability conditions objects with masses as high as
$10^{55}$ g can also be obtained (for a detailed discussion of the
properties of the minimum mass particles see~\cite{BoHa06}).

\section{Bounds on the surface anisotropy of compact objects}

Curvature is described by the tensor field $R^{l}{}_{ijk}$. It is
well known that if one uses singular behavior of the components of
this tensor or its derivatives as a criterion for singularities,
one gets into trouble since the singular behavior of components
could be due to singular behavior of the coordinates or tetrad
basis rather than that of the curvature itself. To avoid this
problem, one should examine the linear and quadratic scalars
formed out of curvature.

In order to find a general restriction for $\Delta (R)$ we shall
consider the behavior of the Ricci invariants
\begin{equation}
r_{0}=R_{i}^{i}=R,
\end{equation}
\begin{equation}
r_{1}=R_{ij}R^{ij},
\end{equation}
and
\begin{equation}
r_{2}=R_{ijkl}R^{ijkl},
\end{equation}
respectively.

If the static line element is regular, satisfying the conditions
$e^{\nu (0)}=\mathrm{constant}\neq 0$ and $e^{\lambda (0)}=1$ ,
then the Ricci invariants are also non-singular functions
throughout the star. In particular for a regular space-time the
invariants are non-vanishing at the origin $r=0$ . For the
invariant $r_{2}$ we find
\begin{eqnarray}  \label{a}
&r_{2}=\left (16\pi \Delta +8\pi \rho +8\pi p_{r}-\frac{4m}{r^{3}}-\frac{%
16\pi \Lambda }{3}\right )^{2}+2\left (8\pi p_{r}-\frac{16\pi \Lambda }{3}+%
\frac{2m}{r^{3}}\right )^{2}+  \notag \\
&2\left (8\pi \rho +\frac{16\pi \Lambda
}{3}-\frac{2m}{r^{3}}\right )^{2}+4\left
(\frac{2m}{r^{3}}+\frac{8\pi \Lambda }{3}\right )^{2}.
\end{eqnarray}

For a monotonically decreasing and regular anisotropy parameter
$\Delta $, the function $r_{2}$ is also regular and monotonically
decreasing throughout the star. Therefore it satisfies the
condition $r_{2}(R)<r_{2}(0),$ leading to the following general
constraint on $\Delta (R)$:
\begin{align}
\Delta (R)&\leq \frac{\langle \rho \rangle +\Lambda
}{3}-\frac{\rho _s}{3}+
\notag \\
\frac{\rho _c}{6}&\sqrt{18\frac{p_c^2}{\rho _c^2}+15+12\frac{p_c}{\rho _c}%
\left (1-\frac{2\Lambda }{\rho _c}\right )+\left (3+2\frac{\langle
\rho
\rangle }{\rho _c}-6\frac{\rho _s}{\rho _c}\right )\frac{\Lambda }{\rho _c}%
-2\left (9\frac{\rho _s^2}{\rho _c^2}-6\frac{\langle \rho \rangle }{\rho _c}%
\frac{\rho _s}{\rho _c}+4\frac{\langle \rho \rangle ^2}{\rho
_c^2}\right )}, \label{b}
\end{align}
where $\rho _{s}$ is the value of the density at the surface of the star, $%
\rho _{s}=\rho (R)$.

Another condition on $\Delta (R)$ can be obtained from the study
of the scalar
\begin{equation}  \label{e}
r_{1}=64\pi ^{2}\left [\left (\rho +\Lambda \right )^{2}+3\left
(p_{r}-\Lambda \right )^{2}+2\Delta \left (\Delta +2p_{r}-2\Lambda
\right )\right ].
\end{equation}

Under the same assumptions of regularity and monotonicity for the functions $%
r_{1}$ and $\Delta (r)$ and considering a non-vanishing surface density $%
\rho _{s}\neq 0$ we find for anisotropy parameter at the surface
of the star the upper limit
\begin{equation}  \label{f}
\Delta (R)\leq \rho _c\sqrt{\frac{1}{2}\left
(1+3\frac{p_{c}^{2}}{\rho
_{c}^{2}}-\frac{\rho _s^2}{\rho _c^2}\right )-\left (3\frac{p_c}{\rho _c}+%
\frac{\rho _s}{\rho _c}-1\right )\frac{\Lambda }{\rho _c}+\frac{\Lambda ^2}{%
\rho _c^2}}+\Lambda .
\end{equation}

The invariant
\begin{equation}
r_{0}=-8\pi \left (\rho -3p_r-2\Delta +4\Lambda \right ),
\end{equation}
leads to the following bound for the anisotropy:
\begin{equation}  \label{bound1}
\Delta (R)\leq \frac{\rho _{c}}{2}\left [3\frac{p_{c}}{\rho
_{c}}+\frac{\rho _{s}}{\rho _{c}}-1\right ],
\end{equation}
which is absolute in the sense that it does not depend on the
value of the cosmological constant.

In the case of isotropic ($\Delta =0$) and stable regular fluid
spheres the condition of monotonic decrease of the scalar $r_{2}$
is always satisfied. By assuming that $\rho _{s}=0$ (a condition
which, for example, is readily satisfied by polytropic equations
of state) from Eq.~(\ref{b}) we obtain the following upper bound
for the mean density of the isotropic star:
\begin{equation}  \label{c}
\langle \rho \rangle \leq \rho _c\sqrt{1+\frac{1}{4}\left (1+3\frac{p_c}{%
\rho _c}\right )^2+\left (1-3\frac{p_c}{\rho _c}\right
)\frac{\Lambda }{\rho _c}+\frac{3}{2}\left (\frac{\Lambda }{\rho
_c}\right )^2}.
\end{equation}

If the central pressure of the star satisfies an equation of state
of the
form $p_c=\rho _c$, and in the absence of the cosmological constant ($%
\Lambda =0$), we obtain the following upper bound for the mean
density of the star:
\begin{equation}
\langle \rho \rangle \leq \sqrt{5}\rho _c.
\end{equation}

For a radiation-like equation of state at the center, $p_c=\rho
_c/3$, the mean density of the star must satisfy the constraint
\begin{equation}
\langle \rho \rangle \leq \sqrt{2}\rho _c.
\end{equation}

These constraints are physically justified since we have assumed a
monotonically decreasing density inside the compact general
relativistic object. The conditions on the anisotropy and mean
density obtained here have been derived only from the study of the
behavior of the curvature invariants, without explicitly solving
the gravitational field equations.

\section{Discussions and final remarks}

The existence of a limiting value of the mass-radius ratio leads
to upper bounds for other physical quantities of observational
interest. One of these quantities is the surface red shift $z$,
defined according to
\begin{equation}
z=\left [1-\frac{2\alpha \left (R\right )M}{R}\right ]^{-1/2}-1.
\end{equation}

In the isotropic case $\Delta =0$ and in the absence of the
cosmological
constant, $\Lambda =0$, Eq.~(\ref{21}) leads to the well-known constraint $%
z\leq 2$ \cite{3,14}. For an anisotropic star in the presence of a
cosmological constant the surface red shift must obey the general
restriction
\begin{equation}  \label{46}
z\leq \frac{2+3f\left (M,R,\Lambda ,\Delta \right )+2\Lambda
/\langle \rho \rangle }{1-2\Lambda /\langle \rho \rangle }.
\end{equation}

By keeping only the first order terms in $\Delta $ and $\Lambda $ Eq.~(\ref%
{46}) can be written as
\begin{equation}
z\leq 2+3f\left (M,R,\Lambda ,\Delta \right )+6\frac{\Lambda
}{\langle \rho \rangle }.
\end{equation}

Therefore much higher surface red shifts than $2$ could be
observational criteria indicating the presence of anisotropic
ultra-compact matter distributions.

By taking into account that the function $\arcsin x/x$ reaches its
maximum value at $x=1$, it follows that the maximum value $f_{\max
}$ of the function $f$ can be approximated as $f_{\max }\approx
\Delta(R)/\langle \rho  \rangle $. Therefore we obtain the
following absolute upper bound for the redshift
\begin{equation}
z\leq 2+\frac{3}{\langle \rho \rangle }\left (\Delta (R)+2\Lambda
\right ).
\end{equation}

With the use of Eq.~(\ref{bound1}) we obtain the following general
restriction on the redshift of anisotropic stars:
\begin{equation}
z\leq 2+\frac{3}{2}\frac{\rho _{c}}{\langle \rho \rangle }\left
(3\frac{p_{c}}{\rho _{c}}+\frac{\rho _{s}}{\rho _{c}}-1\right )+6\frac{%
\Lambda }{\langle \rho \rangle }.
\end{equation}

For high density compact general relativistic objects the mean
density can be approximated by the central density. Thus we have
$\rho _{c}\approx \langle \rho \rangle $. In the limit of high
densities the
equation of state of dense matter satisfies the Zeldovich equation of state $%
p=\rho $. We choose to assume that matter actually behaves in this
manner at densities above about ten times nuclear, that is at
densities greater than $10^{17}$ g/cm$^3$, or temperatures
$T>(\rho /\sigma)^{1/4}$, where $\sigma $ is the radiation
constant \cite{14}. Therefore, by neglecting the surface density
($\rho _{s}/\rho _{c}\approx 0$) and the effect of the
cosmological constant, it follows that the maximum redshift of
anisotropic stars must satisfy the condition
\begin{equation}
z\leq 5,
\end{equation}
value which is consistent with the bound $z\leq 5.211$ obtained by
Ivanov \cite{Iv02}.

However, if the equation of state of the compact matter at the
center of the star satisfies a radiation-type equation of state,
$p=\rho /3$, the redshift of the anisotropic stars satisfies the
same upper bound as the isotropic general relativistic objects,
$z\leq 2$. Therefore the surface redshift is strongly dependent on
the physical conditions at the center of the star. By assuming
that the density of the star is slowly varying, so that
$\rho_c\approx \langle \rho \rangle \approx \rho _s$ and
furthermore the equation of state of the dense matter at the
center of the star satisfies the stiff Zeldovich equation of state
$p_c=\rho _c$, then the surface redshift of anisotropic stars is
constrained by $z\leq 7.5$. Hence very large values of the
redshift may be the main observational signature of anisotropic
stars.

As another application of the obtained upper mass-radius ratios we
shall derive an explicit limit for the total energy of the compact
general relativistic star. The total energy (including the
gravitational field contribution) inside an equipotential surface
$S$ can be defined to be~\cite{15}
\begin{equation}
E=E_{M}+E_{F}=\frac{1}{8\pi }\xi _{s}\int_{S}\left[ K\right] dS,
\label{47}
\end{equation}%
where $\xi ^{i}$ is a Killing field of time translation, $\xi
_{s}$ its value at $S$ and $\left[ K\right] $ is the jump across
the shell of the trace of the extrinsic curvature of $S$,
considered as embedded in the
2-space $t=\mathrm{constant}$. $E_{M}=\int_{S}T_{i}^{k}\xi ^{i}\sqrt{-g}%
dS_{k}$ and $E_{F}$ are the energy of the matter and of the
gravitational field, respectively. This definition is manifestly
coordinate invariant. In
the case of a static spherically symmetric matter distribution from Eq.~(\ref%
{47}) we obtain the following exact expression \cite{15}:
\begin{equation}
E=-re^{\nu /2}\left[ e^{-\lambda /2}\right] .
\end{equation}

Hence the total energy of a compact general relativistic object is
\begin{equation}
E=R\left (1-\frac{2M}{R}-\frac{8\pi }{3}\Lambda R^2\right
)^{1/2}\left [1-\left (1-\frac{2M}{R}-\frac{8\pi }{3}\Lambda
R^2\right )^{1/2}\right ].
\end{equation}

With the use of Eq.~(\ref{21}) we immediately find the following
upper limit for the total energy of the star:
\begin{equation}
E\leq 2R\frac{1+\Lambda /\langle \rho \rangle +3f\left
(M,R,\Lambda ,\Delta \right )/2}{1-2\Lambda /\langle \rho \rangle
} \left (1-\frac{2M}{R}-\frac{8\pi }{3}\Lambda R^{2}\right ).
\end{equation}

For an isotropic matter distribution $\Delta =0$ and
\begin{equation}
E\leq 2R\frac{1+\Lambda /\langle \rho \rangle }{1-2\Lambda
/\langle \rho \rangle }\left (1-\frac{2M}{R}-\frac{8\pi }{3}%
\Lambda R^{2}\right ).
\end{equation}

In the case of a vanishing cosmological constant we obtain the
upper bound
\begin{equation}
E\leq 2R\left (1-\frac{2M}{R}\right ).
\end{equation}

All the previous results on the mass-radius ratio for anisotropic
stellar
objects have been obtained by assuming the basic conditions (\ref{12}) and (%
\ref{16}). But for an arbitrary large anisotropy parameter $\Delta
$ we can not exclude in principle the situation in which these
conditions do not hold. If, for example $\eta (r)>\Psi (r)$
,$\forall r\neq 0$ or
\begin{equation}
\frac{8\pi \Delta (r)}{r}+\frac{d}{dr}\frac{m}{r^{3}}>0,
\end{equation}
then for a star with monotonically decreasing density instead of
the condition (\ref{12}) we must have
\begin{equation}
\frac{d^{2}\Psi }{d\xi ^{2}}>0,\quad\forall r.
\end{equation}

This situation corresponds to a tangential pressure dominated
stellar structure with $\Delta (r)>0$. In this case we obtain a
restriction on the minimum mass-radius ratio of the compact object
of the form
\begin{equation}
\left (1-\frac{8\pi }{3}\Lambda R^{2}\right )\left
[1-\frac{1}{9}\frac{\left
(1-2\Lambda /\langle \rho \rangle \right )^{2}}{\left (1-\frac{%
8\pi }{3}\Lambda R^{2}\right )\left (1+f\right )^{2}}\right
]<\frac{2M}{R}<1.
\end{equation}

For this hypothetically ultra-compact anisotropic star $4/9$ is a
lower bound for the mass-radius ratio.

In the present paper we have considered the mass-radius ratio
bound for anisotropic compact general relativistic objects. Also
in that case it is possible to obtain explicit inequalities
involving $2M/R$ as an explicit function of the anisotropy
parameter $\Delta $. Contrary to the isotropic case we have not
found a universal limit (different from half) for this type of
(possible) astrophysical objects. The surface red shift and the
total energy (including the gravitational one) are strongly
modified due to the presence of anisotropies in the pressure
distribution inside the compact object. The mass-radius ratio
depends very sensitively on the value of the anisotropy parameter
at the surface of the star and different physical models can lead
to very different mass-radius relations. A general feature of the
behavior of physical parameters of anisotropic compact stars is
that the increase in mass, red shift or total energy is
proportional to the deviations from isotropy. Therefore there are
no theoretical restrictions for these stellar type structures to
extend up to the apparent horizon and achieve masses of the order
of $M\leq R/2$ .

\acknowledgments

The authors would like to thank to the two anonymous referees for
comments and suggestions that significantly improved the
manuscript. C.G.B. wishes to thank the Department of Physics of
the Hong Kong University where parts of this work has been
performed. The work of C.G.B. is supported by research grant BO
2530/1-1 of the German Research Foundation (DFG). The work of T.
H. was supported by a Seed Funding Programme for Basic Research of
the Hong Kong Government.

\end{document}